**Body Fat, Skin Tone, and the Accuracy of Smartwatch Caloric Expenditure Estimates**

**Jason Kostrna(corresponding)**

Florida International University, Department of Teaching and Learning, 11200 SW 8th Street, Miami, FL 33199, USA, jkostrna@fiu.edu, (305) 919-4074

https://orcid.org/0000-0002-4838-7094

Ekaterina Oparina

Florida International University, Department of Teaching and Learning, Miami, FL, USA

eoparina@fiu.edu

https://orcid.org/0000-0002-0810-4889

Jessica C. Ramella-Roman

Florida International University, Department of Biomedical Engineering, Miami, FL, USA

jessica.ramella@fiu.edu

https://orcid.org/0000-0002-5710-6004

Cristina Palacios

Florida International University, Department of Dietetics & Nutrition, Miami, FL, USA

cristina.palacios@fiu.edu

https://orcid.org/0000-0001-9437-0376

Andres J. Rodriguez

Florida International University, Department of Biomedical Engineering, Miami, FL, USA

andrrodr@fiu.edu

https://orcid.org/0000-0002-2159-8123

JunZhu Pei

Florida International University, Department of Biomedical Engineering, Miami, FL, USA

jpei@fiu.edu

https://orcid.org/0000-0002-8462-0893

Ajmal Ajmal

Florida International University, Department of Biomedical Engineering, Miami, FL, USA

aajmal@fiu.edu

https://orcid.org/0009-0000-4600-0602

**Competing Interests**

All authors declare no financial or non-financial competing interests.

**Author Contributions**

J.K. conceptualized the study, developed the methodology, supervised all stages of the work, conducted formal analyses, and contributed to data curation, visualization, and manuscript writing, review, and editing.

J.C.R.-R. and C.P. conceived and designed the study, supervised all stages of the work, and contributed to manuscript writing, review, and editing.

E.O. assisted with experimental setup, data collection, investigation, and data curation, and contributed to writing and revising the manuscript.

A.J.R., J.P., and A.A. contributed to the setup, data collection and interpretation, and curation of skin reflectance colorimetry and objective imaging with an experimental Spatial Frequency Domain Spectroscopy system, and assisted in manuscript review, and editing.

**Data availability**

The datasets generated and/or analyzed during the current study are available in the OSF repository, https://osf.io/e745u/?view_only=aabf326dc393434c99f4d37bd17588e8.

**Abstract**

Smart watches are commonly used to provide continuous feedback on activity and caloric expenditure and are leveraged for weight management, clinical decisions, and public health strategies (1, 2). Most wrist-worn wearables combine photoplethysmography, accelerometry, and proprietary algorithms to estimate caloric expenditure (3). Prior work shows substantial error, and potential moderators, skin tone and body fat percentage (BF%), remain underexamined (2). We tested whether brand, BF%, and Fitzpatrick skin type (III–V) predict caloric expenditure error versus indirect calorimetry. Hispanic adults ($n$ = 58) completed a single 10-minute recumbent-cycle protocol with alternating 2-minute intervals at ~64–76% and ~77–95% HRmax (Tanaka formula; 4), bracketed by 5-minute rest/recovery. Participants wore Apple Watch Series 8, Fitbit Sense 2, Samsung Galaxy Watch 5, and Garmin Forerunner 955; COSMED K5 metabolic system provided the criterion. After device-specific data quality filters, analyzable participant–device pairings were Apple = 52, Garmin = 51, Samsung = 50, Fitbit = 44. One-sample tests indicated significant mean bias for three of four devices, $p < .05$. Importantly, the non-significant Fitbit bias depended on device-specific outlier removal. Bias ($M$, $SD$) and 95% CI (kcal): Apple 21.60 (36.63), 11.59–31.60; Garmin 68.61 (55.86), 53.28–83.94; Samsung 56.76 (42.03), 45.11–68.41; Fitbit 3.14 (40.95), −8.96 to 15.24. Mixed-effects models showed a device main effect ($p < .001$), a BF% main effect ($p < .01$), and a device by BF% interaction ($p = .02$): PAEE error increased with adiposity across all brands ($p < .01$). Common smart watches substantially misestimate PAEE relative to indirect calorimetry, with error magnitude increasing as BF% rises and varying by brand. Current consumer devices do not yet provide reliable caloric monitoring for individuals or for research; improving accuracy across body types is essential for clinical and public health applications.

**Body Fat, Skin Tone, and the Accuracy of Smartwatch Caloric Expenditure Estimates**

Wearable fitness trackers have revolutionized personal health monitoring by enabling convenient and continuous estimates of physical activity energy expenditure (PAEE). With over a hundred million users adopting these technologies, their accuracy has become a paramount concern for both research and healthcare (5). Accurate PAEE is central to understanding energy balance, guiding weight management, and informing exercise prescriptions (2). Errors in these estimates can mislead individuals attempting to lose weight, distort clinical recommendations, undermine exercise prescriptions, provide inaccurate information for healthcare decisions, and compromise the validity of population-level research (6, 7). Given the global rise in the prevalence of obesity, with nearly half of the U.S. population projected to have overweight or obesity within the next decade, ensuring that wearables provide reliable PAEE estimates is a pressing scientific and public health concern (8).

Most commercial wearables derive PAEE from heart rate (HR) measured via Photoplethysmography (PPG), the optical method underlying most wrist-worn wearables, often paired with accelerometry and proprietary demographic-based algorithms (3). While PPG offers practical advantages, its reliance on light–tissue interaction makes it highly susceptible to error (7). Skin tone and adiposity are two key factors that can bias PPG-based measurements (2). Melanin in darker skin absorbs more of the green light used by common sensors, attenuating the pulsatile signal and contributing to higher error rates in HR detection (7, 9). Likewise, individuals with higher body mass index (BMI) often exhibit thicker, less perfused skin, which diminishes the intensity of light returning to the sensor (7, 10). Recent modeling studies suggest that obesity may have an even greater impact on PPG signal quality than skin pigmentation. Simulations show up to a 60% reduction in signal amplitude under conditions of severe obesity, compared to a 15% reduction attributable to increased melanin (9). Consistent with these findings, empirical evaluations have shown that wearable HR and energy expenditure readings tend to be less accurate in individuals with darker skin tones or higher BMIs (2). Because these same groups experience disproportionate burdens of obesity and cardiovascular disease, systematic inaccuracies in PAEE estimation have implications for health equity and measurement validity (11).

Physical exercise itself poses additional challenges to the accuracy of wearable devices. PPG measurements become notably less reliable during activity – for example, one study observed roughly 30% higher HR errors during exercise compared to rest (6). Repetitive motion can induce motion artifacts, sometimes causing the device to mistake rhythmic movement for cardiac pulsations (a "signal crossover" effect). Since caloric expenditure rises steeply with intensity, such distortions can bias PAEE estimates precisely when they matter most. Proprietary differences in device algorithms and sensor design introduce another layer of variability, with comparative studies reporting an error of more than 20% in PAEE across widely used models (2). These inconsistencies highlight the need for rigorous and systematic validation of wearables across a broad range of users and conditions.

To date, however, comprehensive evaluations of PPG-based energy expenditure accuracy across devices, skin pigmentation, body composition, and exercise intensities remain scarce. Here, we address this gap by assessing the PAEE estimation accuracy of four popular wrist-worn devices in a demographically varied cohort under controlled exercise conditions. We specifically test the hypothesis that darker skin tones and higher BMI are associated with greater errors in PAEE estimation, particularly during moderate to vigorous exercise, and that these errors vary across different wearable devices. This study aims to clearly delineate the physiological and technical limitations of current wearables and to inform the more equitable and reliable use of PPG technology in health monitoring.

## Method

### Participants

The study's procedures were approved by the Florida International University Institutional Review Board (IRB-22-0471; approved October 12, 2024, recruitment starting date January 9, 2023, completion date September 23, 2025) and conducted in accordance with the Declaration of Helsinki (12). Participants were recruited through flyers placed on campus, targeted departmental email invitations, and through snowball recruitment. To encourage participation, students enrolled in Physical Education: Sports and Fitness and Kinesiology courses were offered extra credit opportunities, with alternative assignments provided for students who chose not to participate. In addition, all participants received a $50 Target gift card as compensation for their time.

Recruitment was stratified by body mass index (BMI) and skin tone to ensure adequate representation across participant groups. Participants were eligible if they were Hispanic between the ages of 18 and 50 years with Fitzpatrick skin types III to V (13). Exclusion criteria included failure to meet these demographic requirements, medical contraindications for exercise as determined by the Physical Activity Readiness Questionnaire (PAR-Q+; 14), or the presence of forearm tattoos that could interfere with optical sensor measurements.

A priori power analysis indicated that 52 participants were required to sufficiently power a multivariate analysis of variance (MANOVA) with two groups and two dependent variables (absolute error and directional error) at $\alpha = .05$, $\beta = .80$, and an effect size of $f^2(v) = .20$ (6). To account for attrition and data loss arising from equipment failure or measurement invalidity, the target sample size was increased to 58 participants. Because early volunteers tended to have a BMI below 30, initial enrollment was weighted toward participants within this lower BMI range. Therefore, recruitment in this category was capped at 33 participants; thereafter, only individuals with a BMI of 30 or greater were recruited until the final target sample of 58 was reached. The final sample consisted of 58 participants (31 females, 53.4%; 27 males, 46.6%), with a mean age of 23 years ($SD = 5.91$). The average BMI for the total sample was 29.73 ($SD = 7.43$). Among participants with BMI < 30 ($n = 33$), the mean BMI was 24.24 ($SD = 3.41$), whereas participants with BMI ≥ 30 ($n = 25$) had a mean BMI of 36.97 ($SD = 4.43$). The distribution of Fitzpatrick skin types was as follows: 51.7% type III ($n = 30$), 41.4% type IV ($n = 24$), and 6.9% type V ($n = 4$).

**Design**

The study employed a mixed factorial design. Participants were randomly assigned to one of several wearable device placement conditions. They completed one exercise bout on a recumbent cycle ergometer following a standardized incremental intensity protocol adapted from the American National Standards Institute Consumer Technology Association testing protocol (15).

**Measures**

Energy expenditure was assessed using a COSMED K5 portable metabolic analyzer (COSMED, Rome, Italy), which provides breath-by-breath gas exchange analysis (16) and is accompanied by a Polar H10 chest strap electrocardiogram (ECG; Polar Electro, Kempele, Finland). Participants simultaneously wore five commercial wrist-worn devices: Apple Watch Series 8 (Apple Inc., Cupertino, CA), Fitbit Sense 2 (Google, Mountain View, CA), Samsung Galaxy Watch 5 (Samsung Electronics, Seoul, South Korea), Garmin Forerunner 955 (Garmin,

Olathe, KS), and Empatica E4 (Empatica Inc., Cambridge, MA). The Empatica E4 (Empatica Inc., Cambridge, MA) measured only heart rate and not PAEE. Device placement on the right and left wrists was randomized and did not have a significant impact on any of the analyses (all $p$ values > .05).

Anthropometric measurements included height, assessed with a stadiometer, weight, measured with a calibrated digital scale, and body fat percentage, determined by both skinfold calipers (17) and handheld bioelectrical impedance analysis (Omron HBF-306C; Omron Healthcare, Kyoto, Japan). Wrist and forearm circumferences were measured with a flexible tape. Skin tone was evaluated using three methods: self-report via the Fitzpatrick Skin Type classification (13), skin reflectance colorimetry with a Nix Pro 2 colorimeter (Nix Sensor Ltd., Hamilton, Ontario, Canada), and objective imaging with an experimental Spatial Frequency Domain Spectroscopy (SFDS) system (18). The colorimeter and the SFDS did not provide reliable estimates and were therefore excluded from analysis. Participants also completed a demographic and medical history questionnaire, along with the PAR-Q+ (14), to confirm eligibility and safety for exercise testing.

**Procedure**

Each laboratory session lasted approximately 60 minutes and was conducted in a controlled environment at Florida International University's kinesiology lab. Upon arrival, participants provided written informed consent and completed demographic, medical history, and PAR-Q+ questionnaires. Anthropometric data, including height, weight, BMI, body fat percentage, and forearm dimensions, were then collected. Skin tone was subsequently assessed using Fitzpatrick classification, a colorimeter, and SFDS imaging.

Following initial assessments, participants were randomly assigned to one of the wearable device placement conditions. All wearable devices were fitted to the wrists, then the Polar H10 chest strap and COSMED K5 analyzer were secured. Participants then completed a 10-minute cycling session on a recumbent ergometer. The bout began with a 5-minute rest period followed by a sequence of alternating intensity intervals. Cycling intensity alternated between moderate (64–76% of maximum heart rate) and vigorous (77–95% of maximum heart rate), with maximum heart rate estimated using the Tanaka formula (HRmax = 208 – 0.7 × age; 4). Each intensity phase lasted two minutes, with the cadence maintained at 85 revolutions per minute, while the experimenter adjusted the resistance to achieve the target heart rate (see Figure 1). At the end of the exercise, a 5-minute rest period was provided to allow the heart rate to return to baseline levels. At the conclusion of the exercise protocol, participants were debriefed, thanked for their time, and compensated. All procedures were preregistered on OSF: https://osf.io/kcbgp/?view_only=be86a25a238247c1a656143617e938aa.

**Figure 1**

*Cycling protocol used during the exercise trial.*

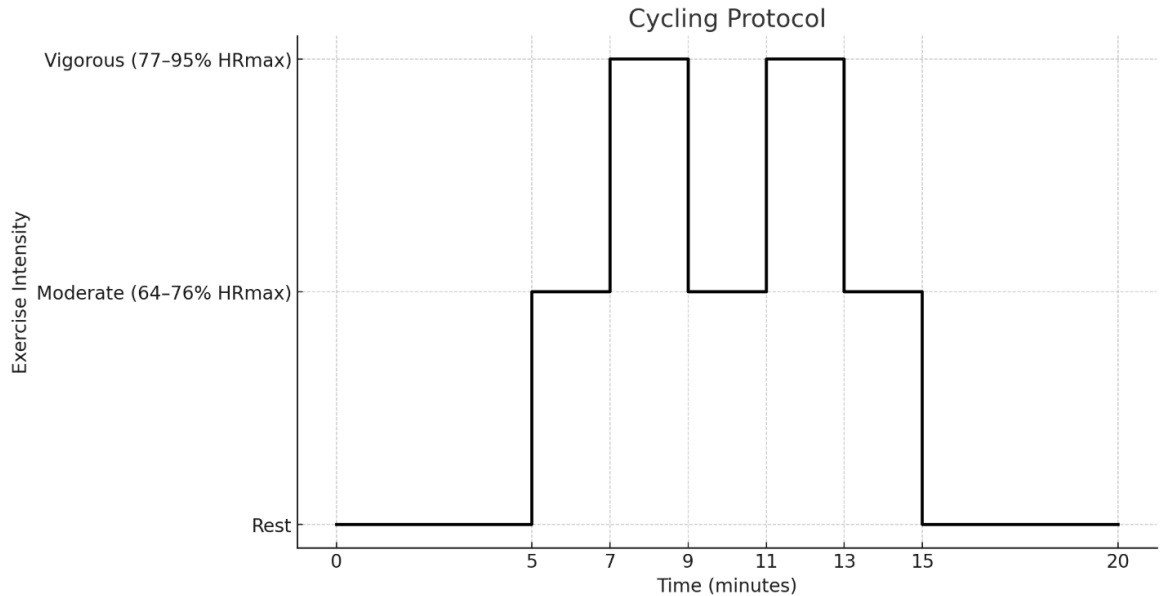

*Note.* After a 5-minute rest period, participants alternated between 2-minute intervals of moderate-intensity cycling (64–76% of their maximum heart rate, HRmax) and vigorous-intensity cycling (77–95% of their maximum heart rate, HRmax). The sequence consisted of three moderate intervals and two vigorous intervals, followed by a 5-minute recovery period.

**Data Analysis**

Device-derived physical activity energy expenditure (PAEE) was evaluated against indirect calorimetry (K5; criterion).

Three complementary error metrics were computed to quantify device accuracy:

Bias = Device − K5

Absolute Error (AE) = |Bias|

Absolute Percentage Error (APE) = 100·AE/K5

Bias represents the directional error in kilocalories (kcal), AE indicates the magnitude of error in kcal, and APE expresses the percentage magnitude error scaled to the criterion measure. Quality control excluded 6 participants due to device errors with the K5 metabolic analyzer. Following the removal of device-specific errors (obvious errors, such as caloric estimates of 0 or 1 calorie or estimates that were over 450% of the K5 estimate), 52 analyzable participant-level pairings remained, with device-specific counts: Apple, 52; Garmin, 51; Samsung, 50; Fitbit, 44. No other participant-level exclusions were applied.

Descriptive statistics were computed by device for each error metric (mean, standard deviation, median, interquartile range, and t-based 95% confidence intervals). For directional error, one-sample t-tests assessed whether the mean bias ≠ was zero within each device. Inferential models tested predictors of error using linear mixed-effects models with $\log(APE + 1)$, $\log(AE + 1)$, and bias (untransformed) as the outcomes, device (4 levels), body fat percentage (BF%; $z$-scored), Fitzpatrick skin type (categorical), and all pairwise interactions as fixed effects; participant was modeled as a random intercept to account for within-person dependence. Type III tests used Satterthwaite approximations (lmerTest). Model fit was summarized with marginal/conditional $R^2$.

Model diagnostics included residual plots, formal tests for heteroscedasticity and normality, and influence checks. Diagnostics indicated homoscedastic error variance but non-normal residuals. To safeguard inference, we supplemented conventional SEs with participant-clustered robust standard errors, which did not alter the significance pattern of primary effects or interactions (clubSandwich CR2). Influence was examined via Cook's distances (influence.ME) and leave-one-participant-out refits, which did not result in changes to the significance outcomes of our models.

For interpretation, we obtained estimated marginal means (EMMs) and simple slopes (emmeans). EMMs were computed at BF% = mean ($z = 0$) and, when appropriate, collapsed across Fitz using the observed sample distribution; for log-models, predictions were back-transformed (exp(pred) − 1) to APE (%) or AE (kcal). Multiple comparisons used Benjamini–Hochberg FDR. Visualization combined violins/boxplots for bias, AE, APE, and model-based curves showing BF% by device effects with 95% confidence intervals. Raw and cleaned data, along with relevant codes for cleaning and analysis, are available on OSF: https://osf.io/e745u/?view_only=aabf326dc393434c99f4d37bd17588e8.

## Results

Across all participants, device estimates showed systematic error relative to indirect calorimetry (K5; COSMED, Rome, Italy). Descriptive analyses indicated that mean bias was significantly different from zero for three of the four devices (one-sample t-tests, all $p < .05$), confirming consistent directional error. Garmin and Samsung systematically overestimated energy expenditure, while Apple's overestimations were noticeably less (Table 1). Fitbit was the only device without a significant mean bias; however, this reflected a misleading average. Fitbit produced seven implausible estimates (> 450% of K5 values), representing approximately 13% of Fitbit trials. When these outliers are included, Fitbit's mean bias rises to 92.3 kcal, by far the largest of any device. Fitbit also failed to generate caloric estimates in one session (Garmin = 1, Samsung = 2, Apple = 0). The absolute error (AE) averaged between ~80–110 kcal, depending on the device, and the absolute percentage error (APE) was consistently high, with median values ranging from ~15–25%. Distributions of bias, AE, and APE are visualized in Figure 2, where violin and boxplots illustrate both magnitude and directionality of device-level errors.

**Table 1**

*Bias, Absolute Error (AE), and Absolute Percentage Error (APE) in Physical Activity Energy*

*Expenditure Estimates Across Four Consumer Wearable Devices*

| | Bias (kcal) | | AE (kcal) | | APE (%) | |
|---|---|---|---|---|---|---|
| Device (*n*) | *M* | *SD* | *M* | *SD* | *M* | *SD* |
| Apple (52) | 21.60 | 36.80 | 33.63 | 26.00 | 35.74 | 27.84 |
| Fitbit† (44) | 3.14 | 40.95 | 31.00 | 26.53 | 31.67 | 27.17 |
| Garmin (51) | 68.61 | 55.86 | 71.47 | 52.07 | 77.56 | 58.23 |
| Samsung (50) | 56.76 | 42.03 | 58.28 | 39.85 | 66.49 | 53.89 |

*Note.* Bias = mean signed error; AE = absolute error; APE = absolute percentage error. †Fitbit

had a substantial number of outlier measurements that were removed. Full device-level

descriptive results are available in the supplemental material (see

PAEE_device_descriptives.xlsx on OSF:

https://osf.io/e745u/?view_only=aabf326dc393434c99f4d37bd17588e8).

**Figure 2**

*Distribution of error metrics in physical activity energy expenditure (PAEE) estimates across four consumer wearable devices (Apple, Garmin, Samsung, Fitbit).*

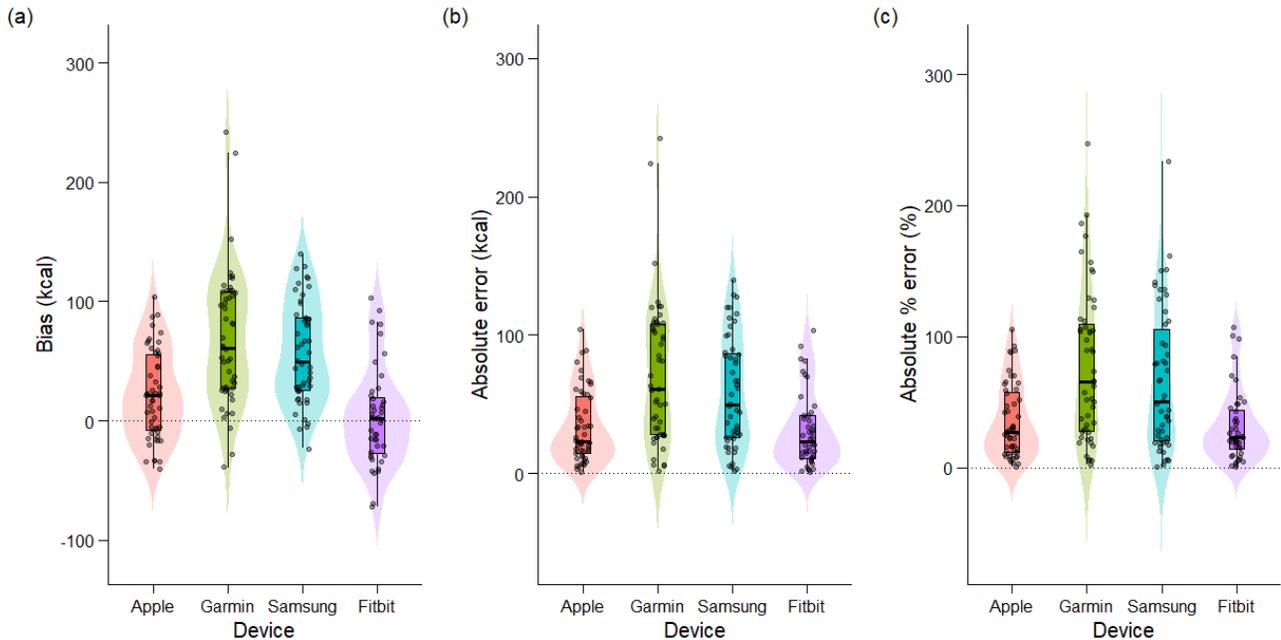

*Note.* Panel (a) shows bias (mean signed error, kcal), panel (b) shows absolute error (AE, kcal), and panel (c) shows absolute percentage error (APE, %). Each violin plot displays the full distribution, boxplots represent interquartile ranges with medians, and dots represent individual participant values.

Mixed-effects models of log-transformed APE revealed significant main effects of device ($p < .001$) and body fat percentage (BF%; $p = .01$), as well as a device × BF% interaction ($p = .02$). Specifically, higher BF% was associated with greater APE across all devices, but slopes differed in magnitude: errors increased more steeply for Fitbit and Garmin compared to Apple (Figure 3). Fitzpatrick skin type showed no robust main effect ($p = .89$). Full model statistics are presented in supplemental data (see PAEE_LMM_results.xlsx on OSF: https://osf.io/e745u/?view_only=aabf326dc393434c99f4d37bd17588e8). Regressions of BF% by device by error are also provided (see PAEE_BF_Regression_Untransformed.xlsx on OSF: https://osf.io/e745u/?view_only=aabf326dc393434c99f4d37bd17588e8).

Mixed-effects models of log-transformed AE revealed significant main effects of device ($p < .001$) and body fat percentage (BF%; $p = .01$), with a significant device by BF% interaction ($p = .02$). At the mean BF%, back-transformed predicted AE was lowest for Apple ($M = 25.15$ kcal, $SE = 4.59$) and Fitbit ($M = 18.82$ kcal, $SE = 3.56$) and highest for Garmin ($M = 47.86$ kcal, $SE = 8.60$) and Samsung ($M = 40.98$ kcal, $SE = 8.42$). As BF% increased, AE rose across all devices, with steeper slopes for Garmin and Samsung relative to Apple (Figure 3). This suggests that the magnitude of error is systematically higher for some devices, particularly at higher adiposity levels. Complete model statistics are presented in supplemental data (see PAEE_AE_Bias_backtrans.xlsx on OSF: https://osf.io/e745u/?view_only=aabf326dc393434c99f4d37bd17588e8).

Bias models showed significant main effects of device ($p < .001$) and BF% ($p < .01$). At the mean body fat percentage, Apple displayed a modest positive bias ($M = 22.77$ kcal, $SE = 7.78$), while Garmin ($M = 37.62$ kcal, $SE = 8.37$) and Samsung ($M = 35.68$ kcal, $SE = 8.46$) significantly overestimated energy expenditure. In contrast, Fitbit demonstrated a negative bias ($M = -19.60$ kcal, $SE = 8.68$), indicating underestimation relative to the criterion after outlier removal. Importantly, BF% was positively associated with bias across devices, suggesting that higher adiposity systematically increased the tendency for devices to deviate from the gold standard (Figure 3). Complete model statistics are presented in supplemental data (see PAEE_AE_Bias_backtrans.xlsx on OSF: https://osf.io/e745u/?view_only=aabf326dc393434c99f4d37bd17588e8).

**Figure 3**

*Associations between body fat percentage (BF%) and physical activity energy expenditure*

*(PAEE) error metrics across four consumer wearable devices (Apple, Garmin, Samsung, Fitbit).*

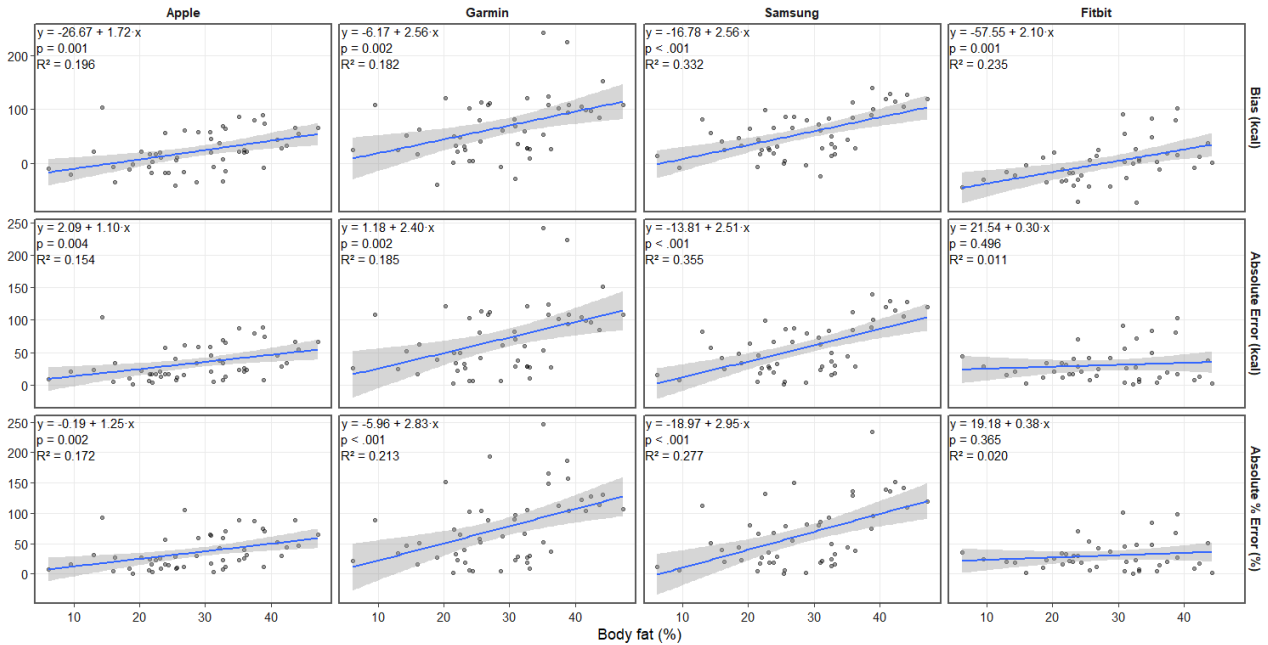

*Note.* Each scatterplot displays regression lines with 95% confidence intervals (shaded areas) for body fat percentage predicting bias (top row), absolute error (middle row), and absolute percentage error (bottom row). Dots represent individual participant values. Regression equations, p-values, and $R^2$ statistics are provided within each panel. Complete descriptive and regression outputs are available in the supplemental material (see

PAEE_AE_Bias_backtrans.xlsx on OSF:

https://osf.io/e745u/?view_only=aabf326dc393434c99f4d37bd17588e8).

## Discussion

This study examined the accuracy of physical activity energy expenditure (PAEE) estimates from four popular wrist-worn wearables (Apple, Garmin, Samsung, and Fitbit) against indirect calorimetry during structured cycling. Three core findings emerged. First, all devices exhibited substantial PAEE error. This result is consistent with prior studies, which report that calorie estimates are inaccurate, with mean absolute percentage errors often exceeding 20% in controlled exercise tasks (2, 19, 20). Furthermore, device-specific differences were evident: Apple produced the smallest bias, whereas Garmin and Samsung systematically overestimated PAEE. Fitbit exhibited the largest variability, with extreme outliers and occasional missing data, which raises concerns about its reliability in research and applied contexts. Second, adiposity consistently moderates error. Participants with higher body fat percentage (BF%) demonstrated greater bias and reduced relative accuracy. These findings align with recent evidence showing that algorithms mapping heart rate and accelerometry to energy expenditure perform less effectively in individuals with elevated adiposity, in part because most models are calibrated on leaner populations and do not account for altered hemodynamics, body composition, or mechanical efficiency in individuals with higher BF% (21, 22). Increased soft tissue motion at the wrist and calibration ranges that underrepresent individuals with high body fat percentage likely compound these errors (9, 23, 24). Third, we found no measurable effect of Fitzpatrick skin tone (III–V) on PAEE error. Although optical heart rate sensors are sensitive to pigmentation (25), the present data align with evidence suggesting that, when tested under standardized conditions, skin tone may exert a smaller or inconsistent influence on PAEE than adiposity (6, 26). This null finding should be interpreted with caution, given the limited sample size, especially regarding Fitzpatrick V, and does not dismiss potential disparities in free-living contexts.

From a user perspective, these results highlight practical risks. Large PAEE errors may mislead individuals attempting to manage their weight or energy balance, with these errors disproportionately affecting those with a higher body fat percentage, a population that is already at high cardiometabolic risk. Clinically and in public health research, uncorrected PAEE bias and outliers can obscure intervention effects and misclassify activity exposures, undermining the validity of epidemiological surveillance and applied studies that rely on wearable-derived metrics. These limitations have already prompted calls for phenotype-specific algorithms and greater transparency in commercial devices (22).

This study offers several strengths, including the use of a structured cycling protocol spanning moderate to vigorous intensities, a gold-standard metabolic criterion (COSMED K5) with documented validity, and evaluation of four major wearable brands. Nonetheless, several limitations should be acknowledged. First, the restriction to a single exercise modality may limit generalizability to free-living movement patterns. In more active modalities, inertial measurement unit (IMU) data can provide additional information for calculating PAEE. Second, the sample consisted of Hispanic adults with Fitzpatrick skin tones III–V, and the small number of participants in the Type V phenotypic subgroup constrained statistical power and may have masked potential effects. Third, because participants only completed a single exercise bout, we were unable to account for potential algorithmic learning effects, whereby devices adapt or recalibrate to users over repeated use, if such adaptations are included in proprietary estimations. Finally, the proprietary nature of device algorithms prevents mechanistic attribution of error patterns and underscores the need for greater transparency from manufacturers.

Given the consistently high PAEE error, particularly among individuals with higher body fat percentages, improvements to the device or algorithm are needed. Future research should include approaches such as training models on datasets stratified by adiposity, sex, and age; enhancing motion-type detection to ensure activity-specific energy expenditure modeling; and integrating hybrid sensing modalities (e.g., multi-wavelength PPG, temperature) to improve accuracy.

**Competing Interests**

All authors declare no financial or non-financial competing interests.

**Author Contributions**

J.K. conceptualized the study, developed the methodology, supervised all stages of the work, conducted formal analyses, and contributed to data curation, visualization, and manuscript writing, review, and editing.

J.C.R.-R. and C.P. conceived and designed the study, supervised all stages of the work, and contributed to manuscript writing, review, and editing.

E.O. assisted with experimental setup, data collection, investigation, and data curation, and contributed to writing and revising the manuscript.

A.J.R., J.P., and A.A. contributed to the setup, data collection and interpretation, and curation of skin reflectance colorimetry and objective imaging with an experimental Spatial Frequency Domain Spectroscopy system, and assisted in manuscript review, and editing.

**Data availability**

The datasets generated and/or analyzed during the current study are available in the OSF repository, https://osf.io/e745u/?view_only=aabf326dc393434c99f4d37bd17588e8.

**Acknowledgements**

The authors acknowledge the invaluable contributions of the undergraduate research team Sabrina Urban, Olivia Roeder, Brooke McCann, Albert Sirven, Alexis Smith, Christina Varela,

Danay Romero, Lorena Velazquez Diaz, Ashley Khouly, Mariana Espinal, and Fabiana Marin Gallucci for their assistance with participant recruitment and data collection. Their enthusiasm, professionalism, and attention to detail were instrumental to the successful completion of this study.

This research was supported by a pilot grant from the National Science Foundation (NSF) via the Precise Advanced Technologies and Health Systems for Underserved Populations (PATHS-UP) program. The funder played no role in study design, data collection, analysis and interpretation of data, or the writing of this manuscript.